\documentclass{aa}

\usepackage{graphicx}
\usepackage{txfonts}
\begin{document} 

   \title{Periodicities of the RV~Tau-type pulsating star DF~Cygni: a combination of \textit{Kepler} data with ground-based observations}
   \titlerunning{The RV~Tau-type pulsating star DF~Cygni}

   \author{A. Bódi
          \inst{1}
          \and
          K. Szatmáry
          \inst{1}
          \and
          L.L. Kiss
          \inst{2,3}
          }

   \institute{Department of Experimental Physics and Astronomical Observatory, University of Szeged, H-6720 Szeged, Dóm tér 9., Hungary
         \and
   Konkoly Observatory, Research Centre for Astronomy and Earth Sciences, Hungarian Academy of Sciences, H-1121 Budapest, Konkoly Thege M. út 15-17, Hungary
         \and
   Sydney Institute for Astronomy, School of Physics A28, University of Sydney, NSW 2006, Australia\\
         \email{abodi@titan.physx.u-szeged.hu}
             }

   \date{Received ...; accepted ...}

 
  \abstract
   {The RV~Tauri stars constitute a small group of classical pulsating stars with some dozen known members in the Milky Way. The light variation is caused predominantly by pulsations, but these alone do not explain the full complexity of the light curves. High quality photometry of RV~Tau-type stars is very rare. DF~Cygni is the only member of this class of stars in the original \textit{ Kepler} field, hence allowing the most accurate photometric investigation of an RV~Tauri star to date.}
   {The main goal is to analyse the periodicities of the RV~Tauri-type star DF~Cygni by combining four years of high-quality \textit {Kepler} photometry with almost half a century of visual data collected by the American Association of Variable Star Observers.}
   {\textit{Kepler} quarters of data have been stitched together to minimize the systematic effects of the space data. The mean levels have been matched with the AAVSO visual data. Both datasets have been submitted to Fourier and wavelet analyses, while the stability of the main pulsations has been studied with the O--C method and the analysis of the time-dependent amplitudes.}
   {DF~Cygni shows a very rich behaviour on all time-scales. The slow variation has a period of 779.606 d and it has been remarkably coherent during the whole time-span of the combined data. On top of the long-term cycles the pulsations appear with a period of 24.925 d (or the double period of 49.85 d if we take the RV Tau-type alternation of the cycles into account). Both types of light variation significantly fluctuate in time, with a constantly changing interplay of amplitude and phase modulations. The long-period change (i.e. the RVb signature) somewhat resembles the Long Secondary Period (LSP) phenomenon of the pulsating red giants, whereas the short-period pulsations are very similar to those of the Cepheid variables. Comparing the pulsation patterns with the latest models of Type-II Cepheids, we found evidence of strong non-linear effects directly observable in the \textit{Kepler} light curve.
   }
   {}


   \maketitle
%

\section{Introduction}
The RV~Tauri stars are evolved low-mass F, G and K-type pulsating supergiants that are located above the Population II Cepheids in the instability strip. Traditionally, 
they are classified into two subgroups. The RVa type is characterised by alternating minima with typical periods longer than 20 days. The RVb stars show an additional long-term variation in the mean brightness, with typical periods of 700-1200 days. The short-period variation is interpreted with fundamental mode pulsation, while the long-term phenomena is commonly interpreted as being caused by periodic obscuration of a binary system by circumbinary dust disk (Lloyd Evans 1985, Pollard et al. 1996, Van Winckel et al. 1999, Fokin 2001, Maas et al. 2002, Gezer et al. 2015).

The RV~Tauri stars have similar luminosities, but higher effective temperatures than Miras. The luminosity function of RV~Tauri stars mainly overlaps with the low-luminosity part of the Mira luminosity function. There are similarities between their observational characteristics. A few Miras show double maxima which is common in RV~Tauri stars (e.g. R~Cen, R~Nor). Some Miras and semiregular variables may exhibit the same (or physically similar) quasi-periodic, long-term mean brightness variations as do RVb stars (e.g. RU~Vir; Willson and Templeton 2009). Also, there is an extensive literature on the so-called Long Secondary Periods of the Asymptotic Giant Branch stars, a phenomenon that still puzzles the researchers (Wood et al. 1999; Takayama, Wood and Ita 2015) and which causes light curves that are surprisingly similar to those of the RVb stars.

Originally, the RV~Tauri stars were classified as Cepheid variable due to the similarities and the poor quality of the light curves. The Type II Cepheids may be divided in groups by period, such that the stars with periods between 1 and 5 days (BL~Her class), 10–20 days (W~Vir class), and greater than 20 days (RV~Tauri class) have different evolutionary histories (Wallerstein, 2002). In each cases the shape of the light curve is nearly sinusoidal, but the RV~Tau-type stars show alternating minima (which means that every second minimum is shallower). The irregularity of the minima grows as the period becomes longer. The light curve of some RV~Tau-type stars randomly switches into a low-amplitude irregular variation then switches back into the previous state (e.g. AC~Her (Kolláth et al. 1998) and R~Sct (Kolláth 1990; Buchler et al. 1996)). The RV~Tauri variables have been placed among the post-AGB stars (Jura 1986),  that are rapidly evolving descendants of stars with initial masses lower than 8 M$_{\odot}$. In the late stages of their evolution  they are crossing the Hertzsprung–Russell diagram (HRD) from the cool asymptotic giant branch (AGB) to the ionizing temperature of the planetary nebula nuclei. During this process they cross the classical instability strip, in which large-amplitude radial oscillations are driven by the $\kappa$-mechanism, while the complexity of the post-AGB variability pattern depends on their location in the HRD relative to the classical instability strip (Kiss et al. 2007).

Several models attempted to reproduce the light variations of RV~Tauri stars. The short period variation can be easily explained by summing two sine curves if their frequencys' ratio is 1:2 and the phase difference is $\pi$/2 (Pollard et al. 1996). Buchler et al. (1996) and Kolláth et al. (1998) established a more sophisticated model to describe the light curve of R~Sct and AC~Her by low-dimensional chaos. Due to the several phenomena that occur in the light curves and spectra of RVb stars, it is difficult to explain the typical long-term variation. A long-term photometric and a spectroscopic survey was done by Pollard et al. (1996, 1997). They revealed that in some RVb stars, the reddest colours occur slightly after long-term light minimum. Furthermore, the light and colour amplitude of the short-term period is smaller during the long-term minima. They found that the equivalent width of the H$\alpha$ emission lines in the spectrum are varying with the phase of the long-term period. They concluded that the damping of the pulsation amplitude is difficult to be explained by the popular model of light variation of RVb stars by a binary system which is periodically obscured by circumstellar or circumbinary dust disc (Percy 1993; Waelkens and Waters 1993; Willson and Templeton 2009). They also concluded that the H$\alpha$ emission is caused by passing shock waves throughout the stellar photosphere. Pollard et al. (2006) proposed a possible dust-eclipse model which arrangement can explain both the photometric and spectroscopic characteristics of the long-term phenomena. Unfortunately, the quality of the measurements did not allow to investigate the pulsation mechanism in details, except the case of R Sct and AC Her.

Gezer et al. (2015) studied the Spectral Energy Distribution (SED) of Galactic RV~Tauri stars based on WISE infrared photometry. The objects with circumstellar disks have near-IR excess in the SED. The light curve of the members of this group shows variable mean magnitude, while there is a clear correlation between disk sources and binarity. Consequently, binarity is connected to the long-term changes of the mean brightness. On the other hand, both Gezer et al. (2015) and Giridhar et al. (2005) investigated the photospheric chemical anomaly called depletion, as a sign of dust-gas separation. Two scenarios were proposed to objects with anomalous abundances: (1) single stars with dust-gas separation in their stellar wind and (2) binary stars where dust-gas separation is present in a circumbinary disk (Giridhar et al. 2005). The latter is consistent with the RVb binary hypothesis. Gezer et al. (2015) found that the presence of a disk seems to be a necessary but not sufficient condition for the depletion process to become efficient.

The subject of this paper, the bright RVb-type variable (V$_{\rm max}\approx 10.5$ mag, V$_{\rm min}\approx 13$ mag) DF~Cygni was discovered by Harwood (1927). The period was found to be 49.4 days between the principal minima. Some years later a long period of 790$\pm$10 days was found (Harwood 1936, 1937). The period of the radial pulsation is about 50 days, the long secondary period is about 775 days (Percy 2006). The spectral and the color variation was investigated by Preston et al. (1963) who found CN bands in the spectrum. Gezer et al. (2015), in their search for disk sources, labelled DF~Cygni as uncertain based on its SED, while the star only appears as marginally depleted, if at all (Giridhar et al. 2005, Gezer et al. 2015). Everything put together, DF~Cygni can be considered as one of the better known RV~Tauri-type variable stars, and the only member of this class in the original Kepler field. 

Here we present a detailed light curve analysis of DF~Cygni by combining about
48 years of visual observations with the ultra-precise space photometry
from \textit{Kepler}, with a time-span of about 4 years. The two sources of the data allow investigations both with very high frequency resolution and with extreme photometric accuracy at least over two cycles of the RVb variability. In Sect. 2 we describe the data sources and the preparation of the sets before the analyses. The details of the light curve analysis are presented in Sect. 3, with the discussion of results in Sect. 4. A brief summary is given in Sect. 5.


\section{Preparation of the light curves}

The light curve data collected by the American Association of Variable Star Observers (AAVSO) were downloaded from the public website of the AAVSO\footnote{\tt http://www.aavso.org}. From the available types of measurements only the longest dataset, corresponding to the almost five decades of visual brightness estimates, was used. In total, there were 5924 individual visual magnitudes from over 110 amateur astronomers, obtained between October 1968 and July 2015. Their typical precision is in the order of $\pm$0.3 mag (Kiss et al. 1999), so that the next step was to calculate 5-day bins. Larger bin size would have smoothed out too much of the short-period variation, while shorter bins would contain too few data points to average out the observational errors. On average, there are 10-15 points per bin, so that the estimated 
accuracy of the binned points is about $\pm$0.1 mag.

The \textit{Kepler} dataset was downloaded from KASOC database, maintained by the Kepler Asteroseismic Science Consortium (KASC). The original data include, among other control parameters, the Barycentric Julian Date, the raw and corrected \textit{Kepler} fluxes and the estimated uncertainties (which were in the order of 70 ppm per single observation). DF~Cygni, as one of the long-cadence (one point per every 29.4 minutes and short gaps between the quarters) KASC targets, was observed by \textit{Kepler} throughout the whole 4+ years (17 quarters) of operations in the original field. Since \textit{Kepler} rolled 90 degrees every quarter of a year, the image of DF~Cygni falled on different CCD chips each quarter and hence systematic shifts occurred from quarter to quarter. Given that the length of a quarter and the dominant variability time-scale is in the same order, there is a great difficulty in distinguishing the quarter-to-quarter variations from the intrinsic stellar variability.

In our study we followed the same approach as Bányai et al. (2013), who used short segments of the data immediately before and after the gaps between subsequent quarters to stitch the quarters by producing the most continuous curve. For that, \textit{Kepler} fluxes were converted to magnitudes and only vertical shifts were allowed when joining the data from quarter to quarter. In a very small number of cases, further tiny corrections had to be done manually to reach the most satisfactory continuity with no apparent jumps.

   \begin{figure}
   \centering
   \includegraphics[width=\hsize]{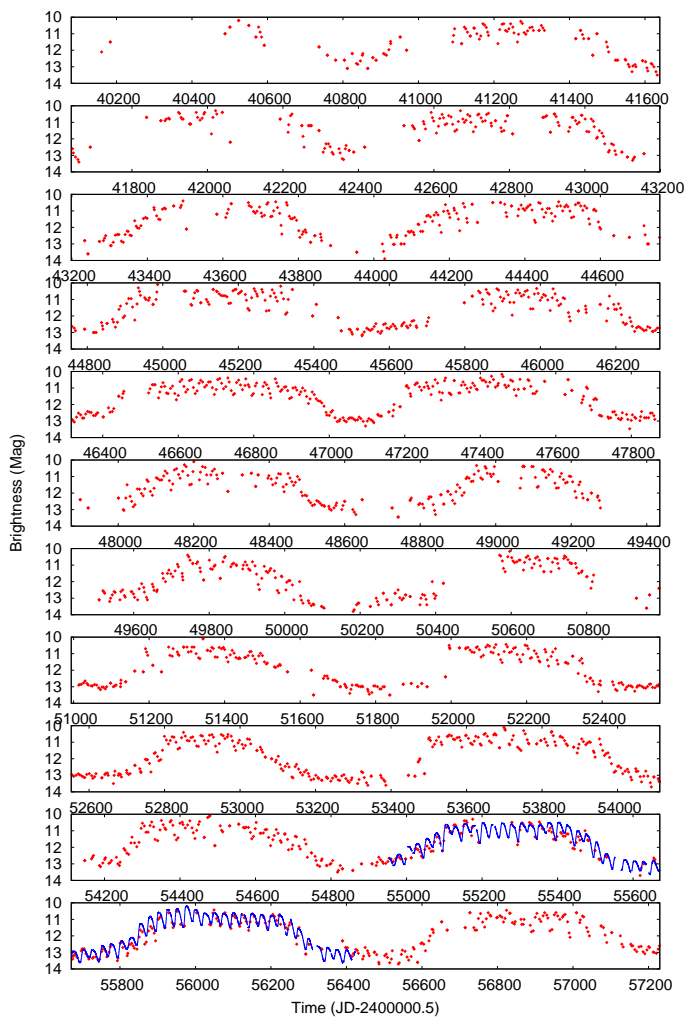}
      \caption{The AAVSO (red dots, 5-day means) and \textit{Kepler} (blue line) light curves of DF~Cygni.}
         \label{Fig:LC}
   \end{figure}

The final light curves of DF~Cygni are plotted in Fig.~\ref{Fig:LC}. Here the small red diamonds show the 5-day means of the AAVSO visual data and the blue line refers to the corrected \textit{Kepler} curve. Note that the zero-point of the stitched \textit{Kepler} data 
was matched with the simultaneous visual light curve, so that the two curves overlap as much as possible. Also note that the subsets shown in the individual panels in Fig.~\ref{Fig:LC} were deliberately selected to cover exactly two cycles of the RVb variability (1570 days=4.3 yr/panel).

It is already apparent from Fig.~\ref{Fig:LC} that the long-period cycles were coherent during the whole 48 years of visual observations. There were slight changes in the shape of the cycles, such as the sometimes flat-shaped maxima, but the faintest states were repeated quite accurately, with no apparent phase shift. On the other hand, the short-period variability changed a lot both in amplitude and the shape of the cycles. The information on the latter is however quite limited given that the 5-days binning results in approximately 10 points per full RV Tau-cycle. The fine details of pulsations are only visible in the continuous \textit{Kepler} light curve. Nevertheless, the two light curves were analysed in a similar fashion with all methods, so that we can directly compare their information content.

\section{Light curve analysis}

The data have been analysed with the traditional methods of Fourier analysis, O--C diagram and wavelet analysis. Some of these were applicable directly to the original data, while some needed further processing steps, like removing the long-term variation of the mean brightness.
All the frequency spectra and then individual frequencies, amplitude and phase values and their uncertainties were calculated with the Period04 software of Lenz \& Breger (2005), using the least squares method when the maximum number of iterations were not reached. The signal-to-noise
ratio for each peak in the frequency spectrum was measured from the mean  local amplitude value in the surrounding region ("noise").
The wavelet map was generated by the Fortran code WWZ of Foster (1996) where the decay parameter "c" was set to 0.0125. This value is appropriate when the time-scale of variability ranges from tens of days to hundreds of days. Details of the time-frequency distribution were extracted with self-developed scripts, which revealed specific information on the stability of the frequency peaks and their amplitudes. 

The AAVSO visual light curve starts from 1968 and now covers 22 long-term cycles. The full peak-to-peak amplitude is about 3.45~mag. Each long-term maxima are flat with similar brightness, but different duration. The minima are mostly shorter and curvy-shaped, showing flat portions with a duration of approximately 100-200 days. The shape of the short-term variation can be hardly recognized in the AAVSO light curve, but the accuracy of the \textit{Kepler} measurement allows us to analyse the 59 short-term cycles covered
from space. The alternation of the minima is clearly visible as well as the variation of the maxima, a behaviour that is largely hidden in the visual dataset.

\begin{figure}
   \centering
   \includegraphics[width=\hsize]{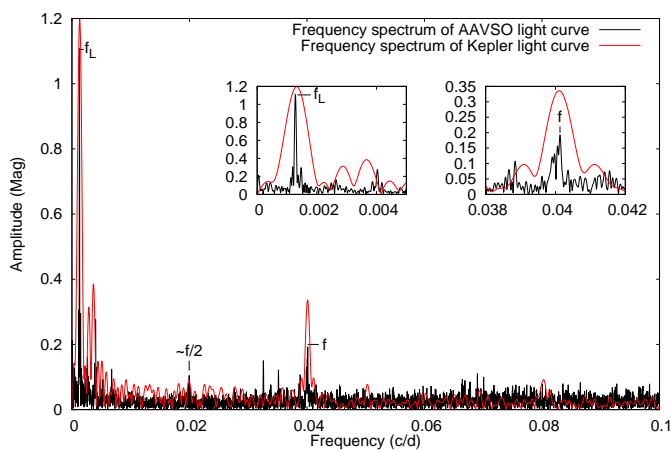}
      \caption{Frequency spectra from the AAVSO (black) and \textit{Kepler} (red) light curves of DF~Cygni. Two zooms to $f_L$ and $f$ are shown in the inserts.}
         \label{Fig:Spectrum}
   \end{figure}

The frequency spectra of the two datasets show two significant peaks corresponding to the long-term (labelled as $f_L$) and the short-term (labelled as $f$) variations (see Fig. \ref{Fig:Spectrum} and Table \ref{table:freqs}).
$f/2$ is also significant in the AAVSO spectrum, but nothing else is above the noise threshold. 
While the widths of the peaks are different, as expected for the different time-span T$_{\rm obs}$ (the frequency resolution scales with 1/T$_{\rm obs}$), it is evident that $f_L$ was much more coherent in the last five decades that $f$. The latter peak at around 0.04~c/d (P$\sim$25 d) is splitted in the AAVSO data, which is caused by the temporal changes of the frequency or the phase. The derived $f_L$ and $f$ values are in good agreement if we compare the AAVSO and \textit{Kepler} data, but the AAVSO set allows a more accurate frequency determination for $f_L$, due to the much longer time-span of the light curve. We note that all the additional peaks in the frequency spectrum of the AAVSO data correspond to the yearly or monthly aliases of three dominant peaks ($f_L$, $f$ and $f/2$). 

\begin{figure}
   \centering
   \includegraphics[width=\hsize]{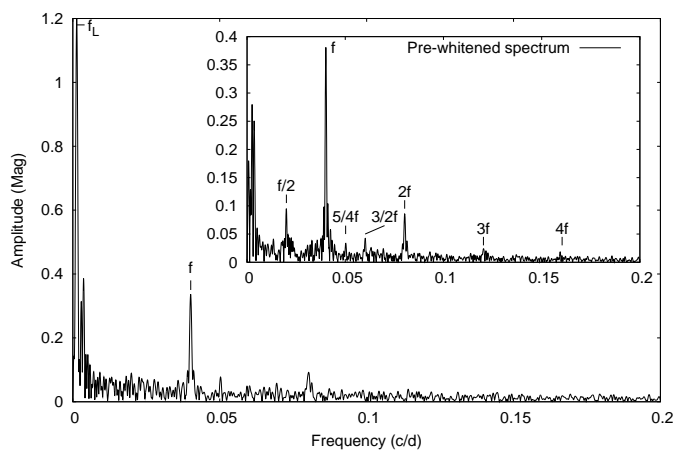}
      \caption{The frequency spectrum of the \textit{Kepler} data. The residual spectrum after pre-whitening with $f_L$ is plotted in the insert.}
         \label{Fig:kepler_spectra}
   \end{figure}

The \textit{Kepler} spectrum displays a significantly richer frequency content. After pre-whitening with $f_L$, the residual spectrum (plotted in the insert of Fig. \ref{Fig:kepler_spectra}) clearly shows the series of integer harmonics ($2f$, $3f$, $4f$) and three extra peaks of subharmonics ($f/2$, $3f/2$ and $5f/4$). There is also excess power in the low-frequency range, which is caused by the unstable shape of the two long-period cycles. Subsequent pre-whitenings resulted in clear detections of further harmonics and subharmonics, up to $6f$ and $7f/2$ (see the full list in Table \ref{table:freqs}). The procedure was stopped when the amplitude of the highest peak (signal) was below the triple value of the mean amplitude around that peak (noise), so that the signal-to-noise ratio was below 3.0. It is well worth noting that no frequency independent of $f_L$ or $f$ was found in the analysis.

\begin{table*}
\caption{Calculated periods, frequencies, amplitudes, phases, their uncertainties and signal-to-noise ratios of the AAVSO and \textit{Kepler} light curves.}             
\label{table:freqs}      
\centering
\begin{tabular}{l c c c c c c c c c}
\hline\hline       
 & ID & P (day) & freq. (c/d) & amp (mag) & $\phi$ (rad/2$\pi$) & $\pm$freq. (c/d) & $\pm$amp (mag) & $\pm\phi$ (rad/2$\pi$) & S/N\\
\hline  

AAVSO & f$_L$ & 779.606 & 0.0012827 & 1.1427 & 0.3710 & 2.57e-07 & 9.09e-03 & 1.27e-03 & 67.48\\
- & f & 24.916 & 0.0401350 & 0.2389 & 0.2138 & 1.23e-06 & 9.09e-03 & 6.06e-03 & 19.03\\
- & $\sim$f/2 & 50.099 & 0.0199604 & 0.0819 & 0.5890 & 3.58e-06 & 9.09e-03 & 1.77e-02 & 5.31\\ \hline
\textit{Kepler} & f$_L$ & 785.824 & 0.00127255 & 1.30205 & 0.417638 & 1.84e-08 & 6.40e-05 & 7.83e-06 & 3580.48\\
- & f & 24.925 & 0.04012110 & 0.39100 & 0.821511 & 6.14e-08 & 6.40e-05 & 2.61e-05 & 870.38\\
- & f/2 & 49.915 & 0.02003410 & 0.08026 & 0.539616 & 2.99e-07 & 6.40e-05 & 1.27e-04 & 211.68\\
- & 2f & 12.460 & 0.08025420 & 0.08660 & 0.593535 & 2.77e-07 & 6.40e-05 & 1.18e-04 & 171.36\\
- & 3/2f & 16.634 & 0.06011710 & 0.04373 & 0.061574 & 5.49e-07 & 6.40e-05 & 2.33e-04 & 107.01\\
- & f/4 & 96.109 & 0.01040490 & 0.03819 & 0.370853 & 6.29e-07 & 6.40e-05 & 2.67e-04 & 108.37\\
- & f/3 & 74.454 & 0.01343110 & 0.02637 & 0.627438 & 9.11e-07 & 6.40e-05 & 3.87e-04 & 74.11\\
- & 3f & 8.308 & 0.12037000 & 0.02669 & 0.243770 & 9.00e-07 & 6.40e-05 & 3.82e-04 & 37.45\\
- & 5/4f & 19.925 & 0.05018830 & 0.01189 & 0.196658 & 2.02e-06 & 6.40e-05 & 8.57e-04 & 26.23\\
- & 4f & 6.235 & 0.16039200 & 0.01090 & 0.162611 & 2.20e-06 & 6.40e-05 & 9.35e-04 & 12.05\\
- & 5/2f & 9.986 & 0.10013900 & 0.01126 & 0.927556 & 2.13e-06 & 6.40e-05 & 9.06e-04 & 19.06\\
- & 7/2f & 7.083 & 0.14118000 & 0.00369 & 0.488105 & 6.51e-06 & 6.40e-05 & 2.76e-03 & 4.85\\
- & 5f & 5.012 & 0.19952900 & 0.00374 & 0.003264 & 6.41e-06 & 6.40e-05 & 2.72e-03 & 4.48\\
- & 6f & 4.172 & 0.23969100 & 0.00199 & 0.033615 & 1.21e-05 & 6.40e-05 & 5.12e-03 & 3.30\\ \hline
\end{tabular}
\end{table*}

\begin{figure}
   \centering
   \includegraphics[width=\hsize]{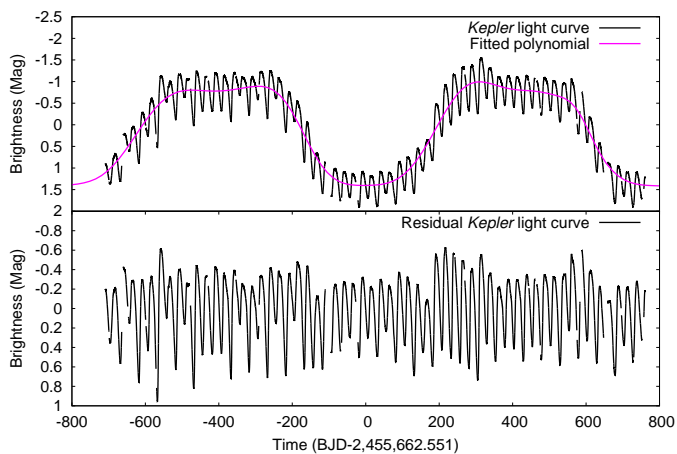}
      \caption{{\it Top:} the original {\it Kepler} light curve (black) and the fitted smooth polynomial (magenta). {\it Bottom:} the residual {\it Kepler} light curve after the subtraction of the long-term variation. The time axis is centered on the long-term minimum.}
         \label{Fig:fitted_polynomial}
   \end{figure}

To analyse the short-period variability separately, we have decomposed the \textit{Kepler} light curve into two components. The smooth, long-term variation was approximated by fitting a polynomial function that was not sensitive to the fast fluctuations due to the pulsations. This has resulted in a somewhat better residual than pre-whitening with the low-frequency sine waves, which were unable to follow the flat-topped maxima of the long-period variation. The components of the \textit{Kepler} light curve are shown in the two panels of Fig. \ref{Fig:fitted_polynomial}, where it is quite apparent that the amplitude of the short-period variability changed seemingly randomly by a factor of 2 during the 4+ years of observations.

\begin{figure}
   \centering
   \includegraphics[width=\hsize]{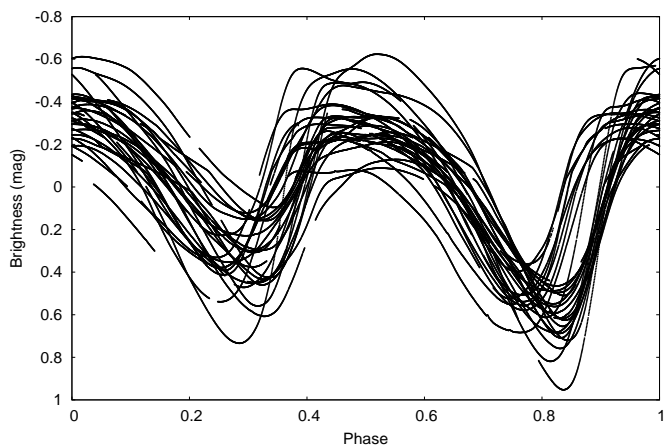}
      \caption{Phase diagram of the residual \textit{Kepler} light curve with a period of 49.99 days after subtraction of the long-term polynomial.}
         \label{Fig:phase_diagram}
   \end{figure}

We plot the phase diagram of the short-term residual variation in Fig.\ \ref{Fig:phase_diagram}. This graph was made using a period of 49.99 days, which is a refined pulsation period from the O--C analysis (see below). While the alternating minima are obvious, it is also visible that the ranges of the maxima and the minima are similarly broad, so that the amplitude variations are not restriced to any special phases of the pulsation cycles. Both ascending branches from the two kinds of minimum are steeper than the descending branches after the consecutive maxima. The overall light curve shape is similar to that of the Cepheid-like pulsators.  

Part of the spread in Fig.\ \ref{Fig:phase_diagram} is caused by variations in phase that can be revealed by the traditional method of the O--C diagram. This is a plot of the differences between the times of observed and calculated minima or maxima as a function of time. This approach is useful when the light curve is long enough and over many cycles the tiny changes accumulate to a detectable shift in phase. For DF~Cygni, the AAVSO data were not suitable for an O--C analysis, because of the scarcity of the 5-day binned light curves, which prevented a detailed analysis of individual pulsation cycles. On the other hand, the accuracy, sampling rate and the continuity of the \textit{Kepler} data allowed an accurate determination of the individual times of minima for each pulsation cycle. We note, however, that a basic assumption of the O--C analysis is the constancy of the light curve shape, i.e. any change in period or phase is just a tiny perturbation to the overall variation. This is not the case for DF~Cygni, but we still feel that there is some value in the method. 

      \begin{figure}
   \centering
   \includegraphics[width=\hsize]{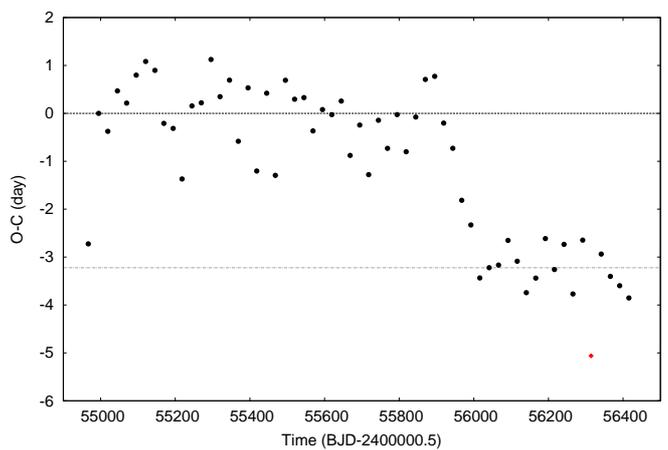}
      \caption{The O--C diagram of the \textit{Kepler} light curve with T$_0$ = 54994.58 and P = 49.99 days. The horizontal lines are the mean values of the group of points.}
         \label{Fig:O--C}
   \end{figure}

To measure the times of minimum we fitted the light curves in a narrow range (1-3 days) of the minima with the following log-normal function:
\begin{equation}
f(t) = \alpha \times \exp{\Bigg(-\ln(2) \times \Bigg(\frac{\ln(2\times\gamma\times\frac{(t-O)}{\beta}+1)}{\gamma}\Bigg)^2\Bigg)} + c
\end{equation}
where $\alpha$, $\beta$, $\gamma$ and $c$ are constant parameters, and $O$ is the observed minimum time that is also obtained by the fitting procedure. The choice of the function was driven by the asymmetry around the minimum and the small number of parameters. The fitted parameters and their uncertainties were calculated using the $\chi^2$ method. The typical uncertainty of $O$ is approximately $\pm$0.0037 d, which corresponds to 5.3 minutes, that is about 1/6th of the \textit{Kepler} long-cadence sampling. As usual in the O--C analyses of RV~Tau-type variables (Percy et al. 1997), we used the epochs of minima because of the sharper shapes around the minimum brightnesses. 

For the O--C plot we adopted the following ephemeris: T${_0}$=BJD 2454994.58 and P=49.99 d. The period value was chosen in such a way that it leads to the flattest O--C plot, predominantly centered on 0. What we found from the \textit{Kepler} data, however, is that the points in the O--C diagram are actually split into two branches with a sharp transition between the two (see Fig.\ \ref{Fig:O--C}). In roughly 100 days between BJD=2455900 and 2456000, there was an apparent phase shift of about 3 days (6\% of the pulsation period), after which the pulsation period remained the same. As it is revealed by the time-frequency analysis, the phase shift coincided exactly with the largest amplitude state of the pulsation, an episode during which the pulsation amplitude was about twice as high than immediately before and after. It is worth noting that the very first point in Fig. \ \ref{Fig:O--C} is a similarly down-scattered value that matches the phase of the data after the phase jump 1000 days later. On the other hand, the cycle-to-cycle scatter is well above the measurement error which is smaller than the symbol size in Fig. \ \ref{Fig:O--C}.

\begin{figure}
   \centering
   \includegraphics[width=\hsize]{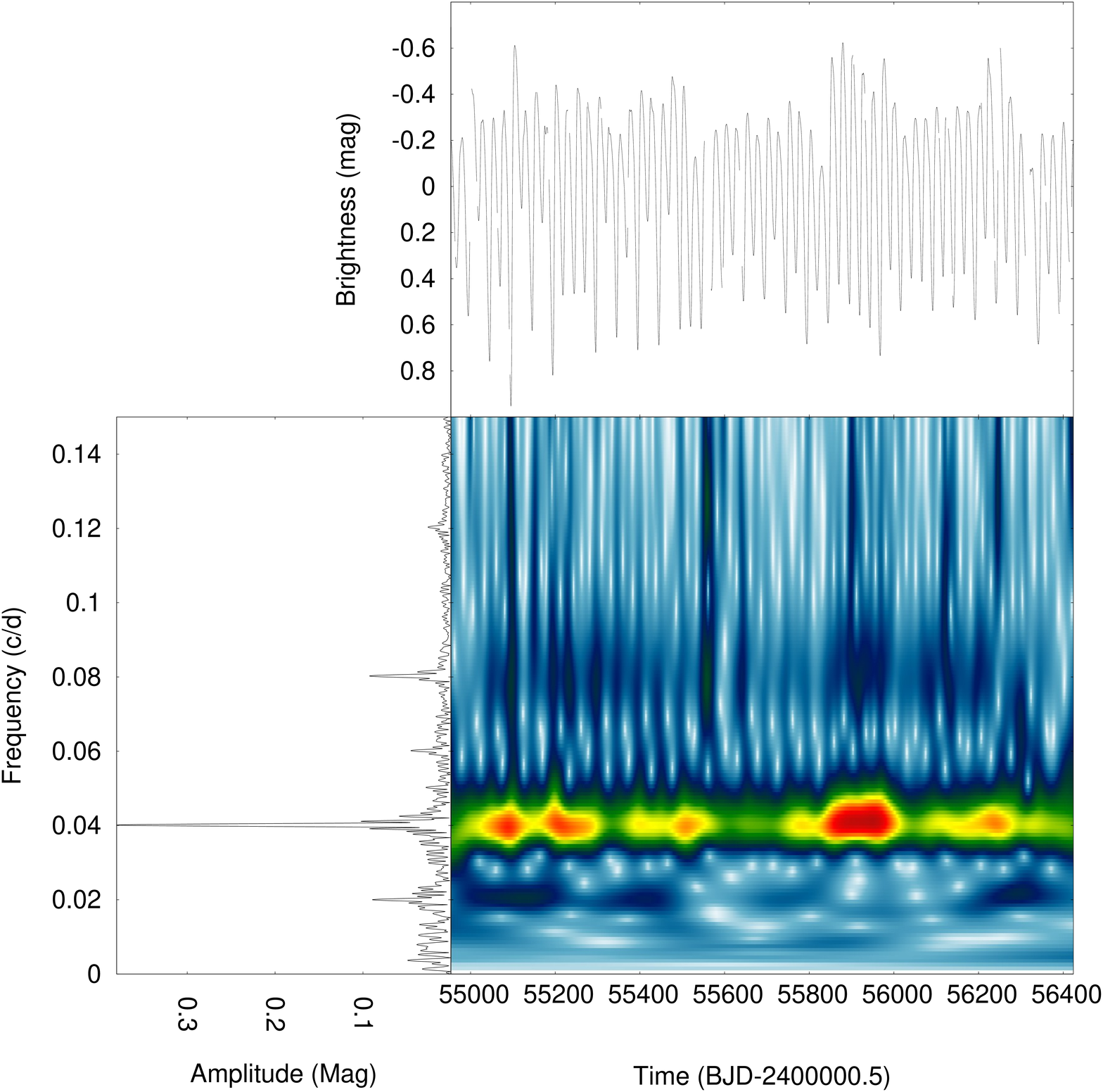}
      \caption{\textit{Top}: the residual light curve after subtracting the slow variability; \textit{bottom}: the frequency spectrum and the wavelet map.}
         \label{Fig:wavelet}
   \end{figure}

To uncover finer details of the changes in amplitude and frequency content as function of time, we calculated the wavelet map, in which the most prominent feature is the amplitude ridge of the frequency $f$ (see Fig.\ \ref{Fig:wavelet}). In addition, the integer and half-integer multiples are also recognizable. For the sake of convenience we plotted the light curve on the top of the map and the Fourier-spectrum on the left-hand side. The colour encoding represents the variations in amplitude, both in time and frequency. It is quite apparent that the amplitudes of all peaks have changed tremendously, whereas the relative differences in the changes of the harmonics indicate significant variations in the overall shape of the light curve.

\begin{figure}
   \centering
   \includegraphics[width=\hsize]{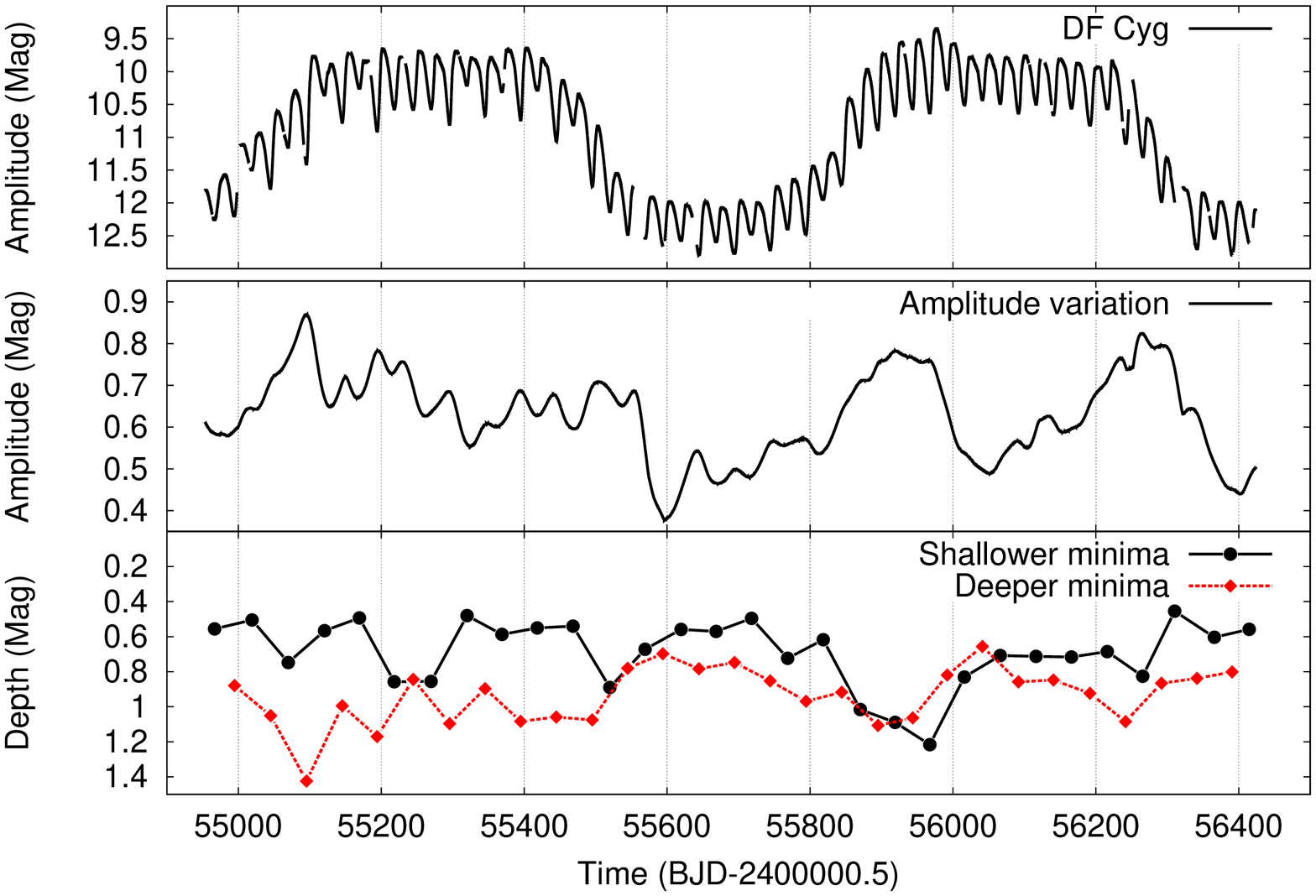}
      \caption{\textit{Top}: the original \textit{Kepler} light curve for comparison. \textit{Middle}: the amplitude variation of the sum of the 0.02, 0.04 and 0.08 c/d frequencies. \textit{Bottom}: the variation of the depth of the shallower (black dots) and deeper minima (red diamonds).}
         \label{Fig:minima_variation}
   \end{figure}

To get a more quantitative picture of the amplitude variations, we have summed the peaks of the ridges of the first three strongest, well-separated frequencies in the wavelet map. For this we traced the peaks of the ridges to measure the instantaneous amplitude of each frequency. Their sum is a quantitave measure of the full oscillation amplitude (note that the role of the phase differences is neglected this way). The result is shown in the middle panel of Fig \ref{Fig:minima_variation}. There seems to be a correlation between the phase of the RVb cycle and the summed amplitude variation. The latter is the smallest when the light curve is in the deep minima (in the middle at BJD 2455600 and at the end, at BJD 2456400). Furthermore, the summed amplitude also displays a sudden decrease right after the second phase jump (at about BJD 2456000). Overall, there is a slight indication that the total amplitude of pulsation is somewhat sensitive to the actual mean brightness, i.e. the pulsation is not entirely coupled from the long-period variation.

We have also examined the variations of the depths of both types of the minima. This was done by subtracting the mean brightness of the preceding and following maxima from the brightness of each minimum. The result is presented in the bottom panel of Fig.~\ref{Fig:minima_variation}. The two kind of depths are plotted separately, with the assumption that the two minima follow a regular alternating pattern. The plot reveals that throughout the whole \textit{Kepler} observations the 'deeper' minimum had indeed almost always a greater depth, except a short transitory phase from BJD 2455900 to 2456000. Interestingly, this behaviour exactly coincides with the $\sim$100 d period when the peak-to-peak amplitude had a maximum and when the phase shift of about 3 days occured.


\section{Discussion}

\begin{figure}
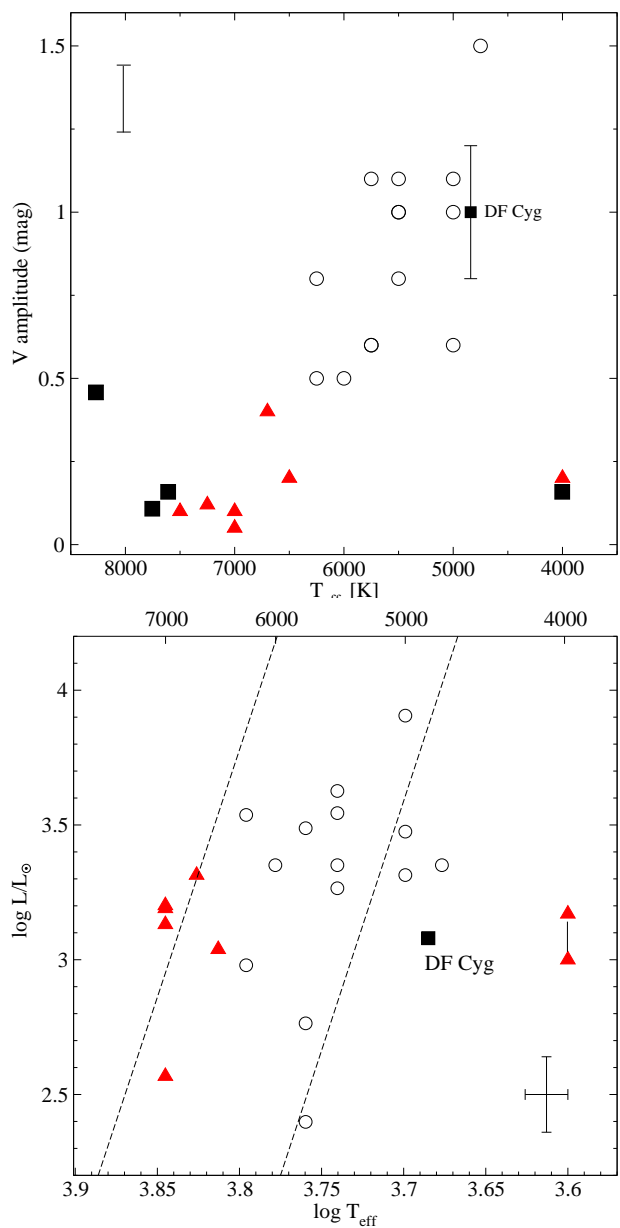

   \centering
   \includegraphics[width=8cm]{teff_amp_df.eps}
   \includegraphics[width=8cm]{hrd_df.eps}
   
      \caption{{\it Top panel:} V-band amplitude as a function of effective 
      temperature for the Kiss et al. (2007) sample of post-AGB variables, supplemented with the
      location of DF~Cygni. Open circles: single periodic stars; triangles: multiperiodic/semiregular stars; squares: variability due to orbital motion. The bar in the upper left corner shows the typical cycle-to-cycle amplitude variation in the pulsating stars.
{\it Bottom panel:} The empirical HRD of the pulsating sample of Kiss et al. (2007) and the location of DF~Cygni. The dashed lines show the edges of the classical instability strip, taken from Christensen-Dalsgaard (2003). The error bars in the lower right corner represent $\pm$3\% error in effective temperature (about 200 K in the range shown) and the $\pm$0.35 mag standard deviation of the LMC P–L relation.}
         \label{hrd}
   \end{figure}

How typical an RV~Tau-type star is DF~Cygni? To shed light on the answer, we plotted DF~Cygni's location in the V-band amplitude vs. effective temperature and the luminosity vs. effective temperature diagrams of selected post-AGB variables studied by Kiss et al. (2007). For the effective temperature we adopted $T_{\rm eff}=$4840 K (Brown et al. 2011; Giridhar et al. 2005), while the V-band amplitude was approximated by the mean peak-to-peak amplitude of the phase diagram in Fig. 5, which is about 1.0$\pm$0.2 mag. The luminosity has been estimated using the RV~Tauri period-luminosity relation in the Large Magellanic Cloud, following the same approach as Kiss et al. (2007). The estimated luminosity is 
$1200\pm300L_\odot$, where the uncertainty reflects the standard deviation of the LMC P-L relation. In the top panel of Fig. 9, DF~Cygni falls close to several high-amplitude stars, all being typical RV~Tau-type objects (such as U~Mon and AI~Sco, both in the RVb class). Similarly good agreement is found in the empirical Hertzsprung-Russell diagram in the bottom panel of Fig. \ref{hrd}. Although DF~Cygni lies somewhat beyond the red edge of the classical instability strip, its location is consistent with its classification. Hence its light curve properties are likely to be representative for the whole class.

DF~Cygni shows a very rich behaviour on all time-scales. The slow variation has a period of 779.606 d and it has been remarkably coherent during the whole 48 years of visual observations. On top of the long-term cycles the pulsations appear with a period of 24.925 d (or the double period of 49.85 d if we take the RV Tau-type alternation of the cycles into account). Both types of light variation significantly fluctuate in time, with a constantly changing interplay of amplitude and phase modulations.

In the Fourier-spectrum of the {\it Kepler} dataset, a characteristic series of subharmonics of the frequency {\it f} appears, which is often interpreted as sign of period-doubling associated with an underlying low-dimensional chaotic behaviour. Similar phenomena were detected in several long-period pulsating variable stars, such as the arch-type of chaotic stars, the RV Tau-type R~Sct (Buchler, Serre \& Koll\'ath 1995, Buchler et al. 1996), the less irregular RVa star AC~Her (Kolláth et al. 1998), several semi-regular variables (Buchler, Kolláth \& Cadmus 2004) and one Mira-type variable (Kiss \& Szatmáry 2002). More recently, \textit{Kepler} has opened a whole new avenue of RR~Lyrae studies based on the period-doubling phenomenon and related effects (e.g. Szabó et al. 2010, Plachy et al. 2013, 2014, Benk\H{o} et al. 2014, Moskalik et al. 2015).   Considering RV~Tau-type variability, there is a strong need for more theoretical investigations. The only recent study that touches the domain in which DF~Cygni resides is that of Smolec (2016), who studied a grid of non-linear convective Type II Cepheid models. Although his study does not cover the full parameter range of RV~Tau-type stars, some of the models extend close to the temperature and luminosity of DF~Cygni. For example, he presented a detailed discussion of the resonances, which affect the shape of the model light curves and in his fig. 13 one can see the regions of the 2:1 resonances of the fundamental, first and second overtone modes. The corresponding Fourier-parameters are shown in figs. 14-15 of Smolec (2016), where DF~Cygni's location is just about outside at the top of the high-luminosity edge of the more massive model calculations. Here, near the red edge of the instability strip, the resonance between the fundamental mode and the first overtone is dominant (see the V-shaped dark features in figs. 13-15, in the rightmost panels with the M=0.8 M$_\odot$ models). The observed complexity of DF~Cygni's pulsations is consistent with the strongly non-linear behaviour of the models. 

As we have seen in Sect. 3, all the periods, amplitudes and phases vary in time, and none of these variations is strictly periodic. As has been revealed by the O--C diagram, there was a transient episode in the light curve around BJD 2456000, where the change of the period and the amplitude was remarkable. Right after the ascending branch of the RVb cycle, the amplitude of the pulsation increased by almost a factor of two, whereas the period decreased quite dramatically (approximately by 3-4 percent, that is about 1 day per cycle). After four to six cycles, the pulsation returned to the previous state. The fact that the sudden amplitude increase is associated with a period decrease is surprising, given that the typical period-amplitude correlation of classical pulsating stars is just the opposite. Interestingly, Smolec (2016) studied the differences between the linear and non-linear pulsation models and found that the relative difference between the non-linear and the linear periods can be up to 15\%, with a strong dependence in the sign across the instability strip. We find a striking feature in fig. 6 of Smolec (2016): models near the location of DF~Cygni show a strong and negative period difference, meaning that the non-linear models have periods that are shorter than those of the linear calculations. We may speculate that it is not a coincidence that DF~Cygni's location and the region where the non-linear pulsation period is significantly smaller agree so well. If the high-amplitude--short-period transient of the light curve is caused by the pulsations becoming more non-linear then the Smolec (2016) models offer a natural explanation to the observed phenomenon. So that the sudden change in the light curve characteristics perhaps can be explained by the emergence of non-linear effects.

Both the variations of the phase diagram and the point-to-point scatter of the O--C diagram (Figs. \ref{Fig:phase_diagram}-\ref{Fig:O--C}) are much unlike the stability of the Cepheid-type variables. We note, however, that recent space observations of Type I Cepheids have found small cycle-to-cycle fluctuations (Derekas et al. 2012), which depend on the mode of pulsation, the first overtone being less stable than the fundamental mode (Evans et al. 2015). Although Type II Cepheids have not yet been covered extensively by space data, one can imagine the outlines of a progression of becoming more unstable on the time-scale of the pulsations as we move closer to the non-linear regime of RV~Tau-type stars.

Besides the properties of the pulsations, the nature of the RVb-phenomenon, the long-term change of the mean brightness, is also worth some discussion. The most common explanation of the RVb-phenomenon is a binary system which is periodically obscured by a circumstellar or a circumbinary dust disc (Waelkens \& Waters 1993, Pollard et al. 1996, 1997, Van Winckel et al. 1999, Maas et al. 2002), although some of the observed characteristics were difficult to reconcile with the dusty disk model (Pollard et al. 1996, 1997). The most recent infrared results from the WISE satellite (Gezer et al. 2015) indicate that RVb stars are exclusively those objects that have well-detected circumstellar disks, while there is also a clear correlation between disk sources and binarity. The 48-years coherence of the long-period variability of DF~Cygni is indeed indicative of a stable mechanisms like binary motion.

We find noteworthy the similarities between the RVb-type variability and the Long Secondary Periods (LSPs) of red giant stars, the latter still representing a mystery (Nicholls et al. 2009 and references therein, Soszynski \& Wood 2013, Saio et al. 2015, Takayama et al. 2015). Among the suggested solutions we find binarity, variable extinction in 
circumstellar disks or strange non-radial oscillations, {many of them} somewhat resembling the proposed explanations of the RVb phenomenon. LSP-like variations have also been found in pulsating red supergiants (Kiss et al. 2006, Yang \& Jiang 2012), suggesting that the phenomenon may not be restricted to the red giants. 
We find some correlations between the pulsations and the RVb-cycle; most notably the amplitude of the pulsations tends to be smaller in the faint states of DF~Cygni, although the case is not entirely clear. It is also interesting to note that  
van Aarle et al. (2011) cross-correlated their sample of candidate post-AGB stars with Long Period Variables (LPV) from MACHO. They found 245 variables falling on the distinct period-luminosity relation of the Long Secondary Periods. While some of those stars 
may not be genuine post-AGB objects, it is an interesting questions if there is some connection between the LSPs of red giants and post-AGB variability. As has been pointed out by the referee, with what we know today, it seems most logical to conjecture that these red giants with long cycles are also binaries, but with longer orbital periods than the RV Tauri stars, so that the same mass-transfer phenomena occurred later in the evolution of the primary, which could climb further on the AGB.


\section{Summary}

The main results of this paper can be summarized as follows:

   \begin{enumerate}
      \item We have combined almost 50 years of visual observations from the AAVSO and about 4 years of \textit{Kepler} data to perform the most detailed light curve analysis of an RV~Tau-type variable star ever obtained. 
      \item The bright RVb-star DF~Cygni is a typical member of the class, showing 
      a prominent long-period mean brightness change and Cepheid-like pulsations that change constantly in amplitude and phase.
      \item The Fourier-spectrum of the \textit{Kepler} data indicate the presence of a 
      complex period-doubling pattern, revealed by the characteristic set of subharmonics. The time-frequency distributions clearly show the non-repetitive 
      variations of the amplitudes.
      \item The \textit{Kepler} light curve also displays some transients, most notably one at BJD 2456000, during which the anti-correlated period and amplitude variations could be an indication of emerging non-linear effects.
      \item The long-term coherence of the RVb modulation is consistent with binary motion, and we note some similarities with the Long Secondary Periods of the pulsating red giants.
   \end{enumerate}

\begin{acknowledgements}
This project has been supported by the Hungarian National Research, Development and Innovation Office (NKFIH) grants K-104607 and K-115709. 
The research leading to these results has received funding from the 
European Community's Seventh Framework Programme (FP7/2007-2013) under 
grant agreement no. 312844 (SPACEINN) and the ESA 
PECS Contract Nos. 4000110889/14/NL/NDe and 4000109997/13/NL/KML. 
We thank the referee, Prof. Christoffel Waelkens for detailed comments that yielded significant improvement to the paper.
LLK thanks the hospitality of the Veszprém Regional Centre of the Hungarian Academy of Sciences (MTA VEAB), where part of this project was carried out. The Kepler Team and the Kepler Guest Observer Office are recognized for helping to make the mission and these data possible. We acknowledge with thanks the variable star observations from the AAVSO International Database contributed by observers worldwide and used in this research. Fruitful discussions with Drs. R. Szabó and E. Plachy are gratefully acknowledged.
\end{acknowledgements}


\end{document}